\documentclass[preprint,aps,12pt,preprintnumbers,nofootinbib,superscriptaddress]{revtex4}

\usepackage{graphicx}
\usepackage{amssymb,amsmath}

\def\Dslash{D\hskip-0.65em /}
\def\dslash{\partial\hskip-0.55em /}

\begin{document}

\newcount\hour \newcount\minute
\hour=\time \divide \hour by 60
\minute=\time
\count99=\hour \multiply \count99 by -60 \advance \minute by \count99
\newcommand{\mydate}{\ \today \ - \number\hour :00}

\preprint{CALT 68-2647}
\preprint{IFT-UAM/CSIC-07-21}
\preprint{UCSD/PTH 07-05}

\title{Neutrino Masses in the  Lee-Wick Standard Model}

\author{Jos\'e Ram\'on Espinosa}
\email[]{jose.espinosa@uam.es}
\affiliation{IFT-UAM/CSIC, Cantoblanco, 28049 Madrid, Spain}

\author{Benjam\'in Grinstein}
\email[]{bgrinstein@ucsd.edu}
\affiliation{Department of Physics, University of California at San Diego, La Jolla, CA 92093, USA}

\author{Donal O'Connell}
\email[]{donal@theory.caltech.edu}
\affiliation{California Institute of Technology, Pasadena, CA 91125, USA}

\author{Mark B. Wise}
\email[]{wise@theory.caltech.edu}
\affiliation{California Institute of Technology, Pasadena, CA 91125, USA}

\date{\today}                                           

\begin{abstract}

Recently, an extension of the standard model based on ideas of Lee and
Wick has been discussed. This theory is free of quadratic divergences
and hence has a Higgs mass that is stable against radiative corrections.
Here, we address the question of whether or not it is possible to couple
very heavy particles, with masses much greater than the weak scale,
to the Lee-Wick standard model degrees of freedom and still preserve
the stability of the weak scale. We show that in the LW-standard model
the familiar see-saw mechanism for generating neutrino masses preserves
the solution to the hierarchy puzzle provided by the higher derivative
terms. The very heavy right handed neutrinos do not destabilize the
Higgs mass. We give an example of new heavy degrees of freedom that would
destabilize the hierarchy, and discuss a general mechanism for coupling
other heavy degrees of freedom to the Higgs doublet while preserving
the hierarchy.

\end{abstract}

\maketitle
\newpage

In a recent paper~\cite{us}, ideas proposed by Lee
and Wick~\cite{Lee:1969fy,Lee:1970iw} were used to extend the standard model so
that it does not contain quadratic divergences in the Higgs mass. Higher
derivative kinetic terms for each of the standard model fields were
added which improve the convergence of Feynman diagrams and give rise
to a theory in which there are no quadratically divergent radiative
corrections to the Higgs mass. The higher derivative terms induce new
poles in the propagators of standard model fields which are interpreted as
massive resonances. These resonances have wrong-sign kinetic terms which
naively give rise to unacceptable instabilities. Lee and Wick propose
altering the energy integrations in the definition of Feynman amplitudes
so that the exponential growth does not occur. It appears that this can
be done order by order in perturbation theory\footnote{This is somewhat
controversial.  See~\cite{Nakanishi:1971jj,Lee:1971ix, Cutkosky:1969fq}.}
in a way which preserves unitarity. However, there is acausal behavior
due to this deformation of the contour of integration. Physically this
acausality is associated with the future boundary condition needed to
forbid the exponentially growing modes. As long as the masses and widths
of the LW-resonances are large enough, this acausality does not manifest
itself on macroscopic scales and is not in conflict with experiment.
The proposal to use Lee-Wick
theory for the Higgs sector of the standard model was first presented
in~\cite{Kuti}.

The massive resonances associated with the higher derivative terms in
Lee-Wick theories have unusual properties. For example, they correspond
to poles on the physical sheet in scattering amplitudes. At the LHC,
we may well discover new resonances, and it would be interesting to
determine whether they are of normal or Lee-Wick type. This issue has
recently been discussed in~\cite{Rizzo:2007ae}.

In the minimal standard model the fermions get their masses through
Yukawa couplings to the Higgs doublet. Gauge invariance forbids
traditional mass terms. These Yukawa couplings do not give mass to
the left handed neutrinos. To describe neutrino masses, one can extend
the particle content to include right handed neutrinos. Right handed
neutrinos have no standard model gauge quantum numbers and so 
Majorana mass terms for them are allowed. If the right handed
neutrino Majorana masses are very large we can understand the
smallness of the observed neutrino masses, since the light
neutrino masses scale as $m_{\nu} \sim v^2/m_R$, where $v$ is the
vacuum expectation value for the Higgs doublet and $m_R$ is the mass
scale associated with the right handed neutrino Majorana masses.
This attractive picture for the generation of neutrino masses is known
as the see-saw mechanism~\cite{Gell-Mann:1980vs}.

Since the generation structure and the quarks are not the focus of
this paper, let us simplify the notation by just considering a single
standard model generation of leptons containing the left handed
doublet denoted by $L$ and the right handed singlet $e_R$. Adding
the right handed neutrino $\nu_R$, the lepton sector of the standard
model has Lagrange density,
\begin{equation}
{\cal L}= {\bar L}i \Dslash L+{\bar e_R}i \Dslash e_R +{\bar \nu_R}i\dslash \nu_R-(m_R \bar \nu_R^c \nu_R+g_e \bar e_R L H^{\dagger} + g_Y \bar \nu_R H^T \epsilon L +{\rm h.c.}).
\end{equation} 

\begin{figure}
\centering
\includegraphics{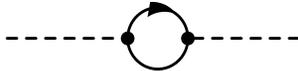}
\caption{One loop correction to the Higgs doublet mass. The dashed line
represent the Higgs scalar, the solid arrowed line is the left handed
lepton while the plain solid line is the right handed neutrino.}
\label{fig:radcorrection}
\end{figure}

It was pointed out in~\cite{Casas:2004gh} that the Feynman diagram in
Fig.~\ref{fig:radcorrection}  gives
a contribution to the mass term for the Higgs doublet that is
quadratically divergent. If one uses dimensional regularization, which
throws away quadratic divergences, there is still a finite correction
\begin{equation}
\delta m_H^2\simeq -{g_Y^2 \over 8 \pi^2}m_R^2 {\rm log}(m_R^2 /\mu^2) 
\end{equation}
which is large compared to the physical mass squared of the Higgs boson if
$m_R \gtrsim 10^7$ GeV~\cite{Casas:2004gh}. This is a manifestation of the
hierarchy problem. In this paper we show that if one used the LW-standard
model this does not occur. Even though the right handed neutrinos are
very heavy the higher derivative kinetic terms for the standard model
fields are powerful enough to prevent the Higgs mass squared from getting
a radiative correction that is proportional to $m_R^2$.

For simplicity, we gauge only $SU(2)_W$ so there is one set of
gauge bosons, $\hat A_{\mu}^A$. The LW-standard model can be
formulated either as a higher derivative theory, or as a theory
without higher derivatives but with auxiliary LW-fields. For the
purposes of the present discussion, it is convenient to work with
the higher derivative version of the theory. To emphasize that this
is the LW-extended model, the fields with higher derivative kinetic
terms are denoted by the presence of a hat. In this simplified
version of the LW standard model the Lagrangian density is,
\begin{eqnarray}
\label{lagrangian}
&&{\cal L}= - \frac{1}{2} {\rm tr} \hat{F}_{\mu \nu} \hat{F}^{\mu\nu} + \frac{1}{M_A^2} 
{\rm tr} \left( \hat D^{\mu} \hat F_{\mu\nu} \right) \left( \hat D^\lambda \hat F_\lambda{}^\nu \right) 
+\left( \hat D_\mu \hat H \right)^\dagger \left(\hat D^\mu \hat H\right) - 
\frac{1}{M_H^2} \left( \hat D_\mu \hat D^\mu \hat H \right)^\dagger \left( \hat D_\nu \hat D^\nu \hat H \right)\nonumber \\
&&- V(\hat H) +\overline{\hat L} i \hat \Dslash \,{\hat L} + \frac{1}{M_L^2} \overline{\hat L} i \hat \Dslash \hat \Dslash \hat \Dslash \, \hat L+ \overline{\hat e}_R i \dslash \,{\hat e}_R+\frac{1}{M_E^2} \overline{\hat e}_R i \dslash \dslash \dslash \, \hat e_R+{\bar \nu_R}i\dslash \nu_R\nonumber \\
&& -(m_R \bar \nu_R^c \nu_R+g_e \overline{\hat e}_R \hat L \hat H^{\dagger} + g_Y \bar \nu_R \hat H^T \epsilon \hat L +{\rm h.c.}).
\end{eqnarray}
Note that we have not added any higher derivative terms for the right
handed neutrino. Calculating the diagram in Fig.~\ref{fig:radcorrection}
in the LW-standard model and using a momentum cutoff $\Lambda$ to
regularize the ultraviolet divergence we find (neglecting the Lee-Wick
mass parameter $M_L$ in
comparison with $m_R$ and $\Lambda$) that,
\begin{equation}
\delta m_{\hat H}^2=-{g_Y^2 \over 8 \pi^2}M_L^2 {\rm log}\left({m_R^2+\Lambda^2 \over m_R^2}\right).
\end{equation}
This leads to acceptably small corrections to the Higgs mass if $g_Y M_L
\lesssim 10$ TeV.\footnote{If we include a higher derivative term for the
right handed neutrino in Eq.~\eqref{lagrangian}, the correction to
the Higgs mass is still proportional to $g_Y M_L$, leading to the same
conclusion.} Thus, we have shown that, at one loop order, the Higgs mass
is not destabilized by the presence of the right handed neutrino. To go
further, we will establish a power counting argument which shows that
the divergence in the Higgs mass squared is at most logarithmic to all
orders of perturbation theory. This is sufficient to show that there
are no large finite corrections to the Higgs mass since we take $m_R$
of order the cutoff in our power counting.

To construct a perturbative power counting argument that shows to
all orders in perturbation theory there is no quadratic divergence
in the Higgs doublet mass term, we must fix a gauge in the higher
derivative theory. We choose to add a covariant gauge fixing term
$- (\partial_\mu \hat{A}^{A \mu})^2 / 2 \xi$ to the Lagrange density
and introduce Faddeev-Popov ghosts that couple to the gauge bosons
in the usual way. Then the propagator for the gauge field is
\begin{equation}
\hat D_{{\mu \nu}}^{AB}(p) = \delta^{AB}\frac{-i}{p^2 - p^4/M_A^2} \left( \eta_{\mu \nu} - (1 - \xi) \frac{p_\mu p_\nu}{p^2} - \xi \frac{p_\mu p_\nu}{M_A^2} \right) .
\end{equation}
We work in Landau gauge, $\xi = 0$, where the gauge boson propagator
scales as $p^{-4}$ at high energy. The propagator for the Higgs
scales at large momenta as $p^{-4}$ while the LW-standard model
leptons, $\hat L$ and $\hat e_R$, have  that scale as $p^{-3}$ at
large momenta. Finally the right handed neutrino propagator and
the Faddeev-Popov ghost propagator scales as $p^{-1}$ and $p^{-2}$,
as usual. There are five kinds of vertices: those where only gauge
bosons interact, vertices where gauge bosons interact with two
scalars, and vertices where two ghosts interact with one gauge
boson. A vertex where $n$ vectors interact (with no scalars) scales
as $p^{6 -n}$, a vertex with two scalars and $n$ vectors scales as
$p^{4 - n}$, while a vertex with two fermions and $n$ vectors scales
as $p^{3 - n}$. The vertex between two ghosts and one gauge field
scales as one power of $p$, as usual, and the vertex from the Yukawa
interaction of the Higgs doublet with the fermions has no factors
of momentum.

Consider an arbitrary Feynman graph with $E$ external Higgs lines,
$L$ loops, $I^\prime$ internal vector lines, $I$ internal scalar
lines, $I_R$ internal right handed neutrino lines, $I_L$ standard
model lepton lines and $I_g$ internal ghost lines, and with
$V_n^\prime$ vector self-interaction vertices, $V_n$ and $\bar V_n$
vertices with $n$ vectors and two scalar Higgs particles or left
handed leptons, respectively.  We also suppose there are $V_g$ ghost
vertices and $V_Y$ Yukawa vertices with two fermions and a Higgs
doublet. Then the superficial degree of divergence, $d$, is
\begin{equation}
d = 4 L - 4 I^\prime - 4 I -I_R-3I_L- 2 I_g + \sum_n V_n^\prime (6-n) + \sum_n V_n (4-n) \sum_n {\bar V}_n (3-n)+ V_g.
\end{equation}
We can simplify this expression using some identities. First, the number
of loops is related to the total number of propagators and vertices by
\begin{equation}
L = I + I^\prime+I_R+I_L +I_g - \sum_n (V^\prime_n + V_n+{\bar V}_n) -V_Y- V_g + 1 ,
\end{equation}
while the total number of lines entering or leaving the vertices is
related to the number of propagators and external lines by
\begin{equation}
\sum_n \left( n V^\prime_n + (n+2) V_n +(n+2) {\bar V}_n \right) + 3 V_g+3V_Y = 2( I+I^\prime +I_R+I_L + I_g) + E,
\end{equation}
where $E$ is the number of external scalars. Finally, we have the
additional relations,
\begin{equation}
2 \sum_n V_n +V_Y= 2 I + E , \;\;\;\; 2 V_g = 2 I_g,~~~~V_Y=2I_R,~~~~\sum_n {\bar V}_n+V_Y=I_R+I_L.
\end{equation}
With these identities in hand, we may express the superficial degree of
divergence as
\begin{equation}
d = 6 - 2 L -V_Y-E .
\end{equation}
Scalar mass renormalizations have $E=2$. The only possible quadratic
divergence in the scalar mass is at one loop with $V_Y=0$. As was
discussed in ~\cite{us}, gauge invariance removes this potential quadratic
divergence. Diagrams involving the leptons have at least $V_Y=2$ and so
are at most logarithmically divergent. Diagrams with other external lines
(which can be subdiagrams in the calculation of the Higgs mass term)
can be analyzed similarly and do not change our conclusions. 

We have shown that in at least one case it is possible to couple
LW-standard model fields to degrees of freedom that are much heavier
and still preserve the stability of the Higgs mass. Furthermore this
case is well motivated by the observed neutrino masses.  However,
this result is not true in general. Suppose, for example, there was
a very heavy complex (normal) scalar $S$. An interaction term of the
type ${\cal L}_{\rm int}=g {\hat H}^{\dagger}{\hat H} S^{\dagger}S$
would lead to a large contribution to the Higgs boson mass. However,
consider coupling the Higgs to a gauge singlet scalar $\hat S$ which
has a higher derivative term in its Lagrange density:
\begin{equation}
\mathcal{L} = \left( \partial_\mu \hat S \right)^\dagger \partial^\mu \hat S - M^2 \hat S^\dagger \hat S
- \frac{1}{m^2} \hat S^\dagger \partial^4 \hat S + g \hat H^\dagger \hat H \hat S^\dagger
\hat S .
\end{equation} 
Then the $\hat S$ propagator is given by
\begin{equation}
\hat D = \frac{m^2}{p^4 - p^2 m^2 + M^2 m^2}.
\end{equation}
If we take the mass parameter $M$ to be large, as in the case of the
scalar $S$, and choose the mass parameter $m$ to be of order of the
weak scale, then the radiative corrections to the Higgs mass are still
small despite the presence of the large scale $M$. The scalar $\hat S$
has unusual properties: for example, from the location of the poles
in its propagator, one can see that it has a tree-level width which is
large compared to its mass. We have not studied the consistency of this
approach in detail.

In summary, we have shown in this paper that it is possible to couple
the Lee-Wick standard model to physics at a much higher scale without
destabilizing the Higgs mass. One of the best motivated examples of high
scale physics is provided by experimental information on neutrino masses,
and we find that the Lee-Wick standard model can easily be extended to
incorporate a heavy right-handed neutrino without reintroducing fine
tuning of the Higgs mass. In addition, we have briefly described a
scenario in which more general physics can be coupled to the Lee-Wick
standard model while maintaining a naturally light Higgs.

\acknowledgements

We thank Andrew Cohen for stimulating discussions. JRE is supported
in part by CICYT, Spain, under contract FPA2004-02015; by a Comunidad
de Madrid project (P-ESP-00346); and by the European Commission under
contracts MRTN-CT-2004-503369 and MRTN-CT-2006-035863. The work of BG
was supported in part by the US Department of Energy under contract
DE-FG03-97ER40546, while the work of DOC and MBW was supported in part
by the US Department of Energy under contract DE-FG03-92ER40701.

\end{document}